\documentclass[aps,groupedaddress,twocolumn]{revtex4}

\usepackage{amsmath}
\usepackage[dvips]{color}
\usepackage{graphicx}

\begin{document}

\title{Near-field light emission from\\ dark-states inverted exciton occupations}
\author{G. Pistone}
\affiliation{Dipartimento di Fisica della Materia e Tecnologie Fisiche
Avanzate,\\Universit\`{a} di Messina\\ Salita Sperone 31, I-98166
Messina, Italy}
\author{ S. Savasta}
\email[Author to whom correspondence should be addressed; electronic mail: ]{salvatore.savasta@unime.it}
\affiliation{Dipartimento di Fisica della Materia e Tecnologie Fisiche
Avanzate,\\Universit\`{a} di Messina\\ Salita Sperone 31, I-98166
Messina, Italy}
\author{O. Di Stefano}
\affiliation{Dipartimento di Fisica della Materia e Tecnologie Fisiche
Avanzate,\\Universit\`{a} di Messina\\ Salita Sperone 31, I-98166
Messina, Italy}
\author{G. Martino}
\affiliation{Dipartimento di Fisica della Materia e Tecnologie Fisiche
Avanzate,\\Universit\`{a} di Messina\\ Salita Sperone 31, I-98166
Messina, Italy}
\author{R. Girlanda}
\affiliation{Dipartimento di Fisica della Materia e Tecnologie Fisiche
Avanzate,\\Universit\`{a} di Messina\\ Salita Sperone 31, I-98166
Messina, Italy}
\author{Stefano Portolan}
\affiliation{Dipartimento di Fisica, Politecnico di Torino, Corso Duca degli Abruzzi 24, 10129 Torino, Italy}

\date{\today}



\begin{abstract}

We theoretically analyze the carrier capture and distribution among
the available energy levels of a symmetric semiconductor quantum dot
under continuous-wave excitation resonant with the barrier energy
levels. At low temperature all the dot level-occupations but one
decrease monotonically with energy. The uncovered exception,
corresponding to the second (dark) energy level, displays a
steady-state carrier density exceeding that of the lowest level more
than a factor two. The root cause is not radiative recombination before
relaxation to  lower energy levels, but at the opposite, carrier trapping due to the symmetry-induced suppression of
radiative recombination. Such a behaviour can be
observed by collection-mode  near-field optical microscopy.

\end{abstract}

\maketitle 

Understanding how electron-hole pairs distribute among the available
energy levels of semiconductor nanostructures after optical
excitation or current injection is crucial for the development of
novel quantum devices \cite{qdev}. One tool frequently used to
explore the energy distribution of electron-hole pairs is
photoluminescence spectroscopy (PL). It  provides a direct
measurement of the optical density of states times the excitonic
population density as a function of energy. Detailed simulations of
Zimmermann {\it et al} \cite{{zimm},{zimm1}} have clarified many
aspects of the intriguing nonequilibrium  dynamics determining the
distribution of excitonic populations among the available energy
levels in quantum wells (QWs) with interface roughness. Already at
temperatures around 40-50 K the individual occupation of states is
close to the Boltzmann distribution since all states can frequently
emit and absorb phonons to reach equilibrium before radiative
emission. At lower temperature more localized states displaying
smaller phonon scattering rates cannot equilibrate and their
occupations deviate from the Boltzmann distribution, displaying a
reduced energy dependence. Far-field PL spectroscopy provide average
measurements over a distribution of different emission sites. The
opportunity for important insight is often lost by the inability to
resolve finer details within this distribution. Scanning near-field
optical microscopy (SNOM) combines the advantages of nanometric
resolution of scanning-probe microscopy with the unique possibility
of characterizing  quantum systems offered by optical spectroscopy.
The SNOM ability to resolve the individual quantum constituents of
semiconductor nanostructures has been widely demonstrated \cite
{{hess},{guest},{matsuda}}. If the spatial near-field resolution
falls below the extension of confined quantum systems, spatially
resolved PL  maps out the spatial probability distribution of the
wave function times the corresponding level occupation
\cite{APLpippo,runge}. The matrix elements governing the
light-matter coupling are a convolution of the quantum states with
the exciting electromagnetic field. This convolution implies that
exciting a direct gap bulk semiconductor with a light field of a
given wavevector resonant with the energy gap results in exciting
excitons (i.e. bound state interband optical transitions) within the
same wavevector. Succeeding in confining the optical excitation to a
very small volume below the diffraction limit, implies the presence
of optical fields with high lateral spatial frequencies, determining
coupling matrix elements that can differ significantly from
far-field ones
\cite{{APLpippo},{mauritz},{APLomar},{jap},{PRBpippo},{JPCMgiovanna}}.
The most striking manifestation of these effects is the breaking of
the usual optical selection rules and the possibility to excite dark
states whose optical excitation is forbidden by symmetry in the far
field. Spatial maps of dark states in semiconductor nanostructures
were simulated for high-resolution SNOM in absorption and two-photon
experiments
\cite{{mauritz},{prbdistef},{APLHohenester},{PRLHohenester}}.
Moreover, dark states  are not able to  emit light in the far-field,
for they generate only fields with high in-plane wavevectors
corresponding to evanescent waves. Hence their lifetimes at low
temperatures result significantly increased with respect to their
bright counterparts.

In this letter we theoretically investigate the impact of the
reduced dark-states relaxation on the distribution of  electron-hole
pairs among the available energy levels after a far-field continuous
wave excitation. We find that this reduced relaxation significantly
throws off balance the PL population dynamics resulting in striking
deviations from equilibrium. In particular we find  that, even at
steady state, the occupation of the dark first excited state of a
symmetric artificial atom can exceed that of the ground state.
Usually larger populations for higher energy levels in  nanosystems
are observed as consequence of bottleneck effects due to  efficient
radiative recombination occurring before relaxation to the lowest
energy-levels. Here the inverted occupations origin from the
opposite effect: carrier trapping due to the presence of
nonradiative states. While the observation of such effects by means
of far-field spectroscopy is prevented (dark states are not able to
emit),  a SNOM tip can collect the evanescent waves generated by the
occupation of dark states. Hence near-field PL  reveals to be an
excellent laboratory to address general questions regarding
nanoscale energy transfer in open quantum systems.

\begin{figure} \begin{center}
\includegraphics[scale=0.5]{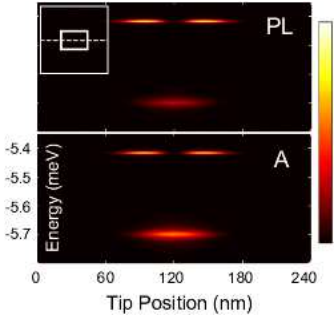}
 \caption{(color online). (a) Near-field PL signal (collection-mode) as a function of photon energy and beam position (line scan shown in the inset) obtained under uniform illumination of the sample at the energy of the QW 1s exciton. T=2 K, FWHM =40 nm.
(b) Total absorption under local illumination, along the same line, spatial resolution and temperature as in (a).}\vspace{-0.4 cm}
 \label{fig1}\end{center}
 \end{figure}
In this work we consider quantum states confined in a naturally
occurring quantum dot, or terrace, induced by monolayer fluctuations
in the thickness of a GaAs/AlGaAs semiconductor quantum well. We
adopt the usual envelope function and assume that the electron-hole
wave functions factorize into a center of mass and a relative part
given by the $1s$ state  of  quantum well excitons \cite{zimm1}. We
assume that the radiative decay of excitons  is not drastically
altered by the presence of the SNOM tip. As pointed out in Ref.
\cite{PRLHohenester}, this is a reasonable assumption since the
photons can be emitted into any solid-angle direction, and the
slightly modified photon density of states in the presence of the
SNOM tip is not expected to be of great importance. Within these
assumptions, recently \cite {APLpippo} it has been shown that the
near-field  spectrally resolved PL signal collected by the tip can
be written as
\begin{equation}
    {I}_{PL}(\bar{\bf R}_{out},\omega_{out}) =  r_0\sum_{\alpha} \left|o^{out}_{\alpha}(\bar{\bf R}_{out})\right|^{2}
    {\cal L}_{\alpha}(\omega_{out})\, N_\alpha \ .
\end{equation}
where $o^{out}_{\alpha}=\int d^{2}{\bf R}\ \tilde{E}_{out} (\bf
R)\psi_{\alpha}\left(\bf R\right)$ is the overlap of the
exciton wavefunctions with the signal mode $\tilde{E}_{out} ({\bf
R})$ supported by the tip centered at the position $\bar{\bf
R}_{out}$ (collection mode), $\pi {\cal
L}_{\alpha}(\omega)=\Gamma/[(\omega-\omega_{\alpha})^{2}+\Gamma^{2}]$, 
where $\Gamma$ is the dephasing rate of the exciton due to radiative
emission and  phonon scattering. 
In Eq. (1) $N_\alpha$ are the diagonal terms of the exciton density matrix.
The kinetic equation for the
diagonal terms of the exciton density matrix  can be derived from
the Heisenberg equations of motion for the exciton operators under
the influence of the system Hamiltonian {\cite {APLpippo}}. The
relevant Hamiltonian is constituted by the bare electronic
Hamiltonian of the semiconductor system, the interaction Hamiltonian
of the semiconductor with the light field and the interaction
Hamiltonian of excitons with the phonon bath. In the following we
consider a low-excitation regime, and according to the dynamics controlled truncation scheme 
\cite{Axt} we include only  states with one electron-hole pair
(excitons). 
 \begin{figure} \begin{center}
\includegraphics[scale=0.5]{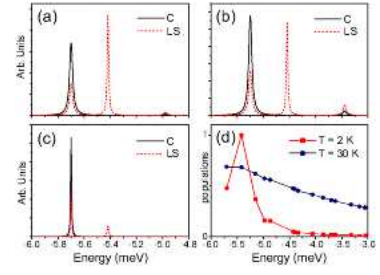}
 \caption{(color online). (a) PL spectra obtained in collection mode ($FWHM =40$ nm and $T=4$ K). The spectra have been obtained centering the tip at one of the two emission maxima of the first excited state (LS) and at the emission maximum of the ground state (C);
(b) Total absorption under local illumination obtained with spatial resolution, at temperature and centering the snom tip as in Fig.\ 2(a);
(c) PL spectra as in Fig.\ 2(a) where the radiative decay rates of all the levels are set two orders of magnitude lower than the lowest nonradiative decay rate.
(d) Level occupations (normalized at the maximum) calculated for two different temperatures.}\vspace{-0.9 cm}
 \label{fig2}\end{center}
 \end{figure}
In addition, we neglect the possible coherent phonon
states and the memory effects induced by the photon and phonon
fields. The resulting set of kinetic equations is {\cite
{zimm,zimm1,APLpippo}}
\begin{equation}
    \partial_t N_{\alpha}=G_{\alpha}\left(\omega_{in} \right)+ \sum_{\beta}\gamma_{\alpha\leftarrow\beta}N_{\beta}-2 \Gamma_\alpha
N_{\alpha}\, ,
\end{equation}
where $\gamma_{\beta\leftarrow\alpha}$ are the phonon-assisted
scattering rates. $2 \Gamma_\alpha = r_{\alpha}+
\sum_{\beta}\gamma_{\beta\leftarrow\alpha}$ is the total
out-scattering rate, being $r_{\alpha}$ the rate for spontaneous
emission, proportional to the exciton oscillator strength: $
r_\alpha = r_0 \left| \int d^{2}{\bf R}\, \psi_{\alpha}\left(\bf
R\right) \right|^2$. In this equation the generation term depends on
the spatial overlap between the illuminating beam and the exciton
wavefunctions: $G_{\alpha}=r_{0}\left|o^{in}_{\alpha}\right|^{2}
{\cal L}_{\alpha}(\omega_{in})$ and $o^{in}_{\alpha}$, analogously
to $o^{out}_{\alpha}$, contains the overlap of the exciton
wavefunctions with the signal mode $\tilde{E}_{in}({\bf R})$
delivered by the tip centered at $\bar{\bf R}_{in}$. In collection
mode (far-field illumination)  the input field can be considered as
uniform  over the QW plane: $\tilde{E}_{in}({\bf
R})=\tilde{E}_{in}^0$.

\begin{figure} \begin{center}
\includegraphics[scale=0.5]{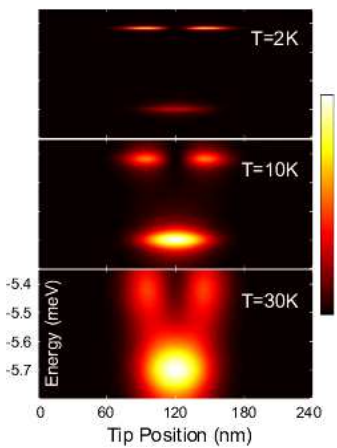}
 \caption{(color online). ((color online). PL spectral line-scans calculated at three different temperatures, with FWHM =40 nm.}\vspace{-1 cm}
 \label{fig3}\end{center}
 \end{figure}
We apply the above developed theoretical scheme to calculate the
individual occupations of exciton states confined in the dot after a
continuous-wave far-field optical excitation resonant with the
energy of the  $1s$ QW exciton (the dot barrier). The obtained
occupations are then used to study the spatially and spectrally
resolved (collection mode) light emission from the dot. The
effective potential felt by 1s excitons used in our simulations is
represented by a sample of $\left(240\times240\right)$ nm with a
prototypical interface-fluctuation confinement of rectangular shape
with dimensions $\left(60\times90\right)$ nm, and monolayer
fluctuations giving rise to a $6$ meV effective confinement
potential. The upper panel of Fig.\ 1 displays  the near-field PL
signal as a function of photon energy and beam position obtained
after uniform illumination of the sample at the energy of the 1s QW 
exciton in the absence of interface fluctuations, fixed as zero of 
energy ($\omega_I =0$ meV). The specific line scan is indicated in
the inset. A Gaussian profile with FWHM = 40 nm of the
electromagnetic-mode supported by the collecting tip has been
assumed. The line scan clearly evidences the first excited state of
the dot which is dark under far-field collection. We observe  that
its spectral line is more intense and narrow (owing to the absence
of radiative decay) than that of the ground state. The calculated PL
spectra shows that dark-states can be observed by high-resolution
SNOM in the usual collection mode configuration after nonresonant
far-field excitation, without the need of nonlinear optical
interactions. Analogous results are expected for locally collected
electroluminescence. For comparison we plotted in the lower panel
the total absorption under local  illumination (FWHM = 40 nm). As
near-field PL is proportional to this quantity times the level
occupations, the comparison provides interesting information about
the steady-state exciton populations. The two panels clearly
indicate that the level occupation of the dark-state  is
significantly larger than that of the ground state, at the opposite
to what predicted by the Boltzmann distribution. The observed
inverted occupations origin from  symmetry-suppression of radiative
decay of the dark-state and the quite small nonradiative scattering
at low temperature ($T= 2 K$) for states well confined in the dot.
This behaviour is better evidenced in panels 2(a) and 2(b). The
spectra have been obtained centering the tip at one of the two
emission maxima of the first excited state (LS) and at the emission
maximum of the ground state (C). In order to better specify the
origin of the inverted occupations, we calculated the PL spectra as
in Fig.\ 2(a) except that we artificially set the radiative decay
rates of all the levels two orders of magnitude lower than the
lowest nonradiative decay rate.  Panel 2(c) displays the results
showing that, as expected  in this case, equilibrium is recovered
and the near-field emission from the dot lowest energy level
dominates even when the tip is centered on the maximum of the
dark-state emission. Fig.\ 2(d) shows the calculated  level
occupations at $T= 2 K$ and $T= 30 K$. At $T= 2  K$ all the
occupations with the striking exception of the second energy level
decrease monotonically with energy. This level displaying an
occupation density which is more then a factor two that of the
lowest energy level. At $T= 30 K$ the monotonous behaviour is
recovered for all the energy levels. We notice that the sum of the
occupations obtained at $T= 30 K$ is significantly larger then that
at $T= 2 K$. At low temperature a large fraction of the resonantly
generated QW excitons (with a quite large radiative decay rate)
recombine by radiative emission, while at higher temperature phonon
scattering lowers  this effect, increasing the carrier density
captured by the dot. Fig.\ 3 shows the near-field ($FWHM =40$ nm) PL
spectral line-scans  obtained at three different temperatures.
Increasing the temperature all states are able to better thermalize,
frequently emitting and absorbing phonons before radiative emission.
\begin{figure} \begin{center}
\includegraphics[scale=0.5]{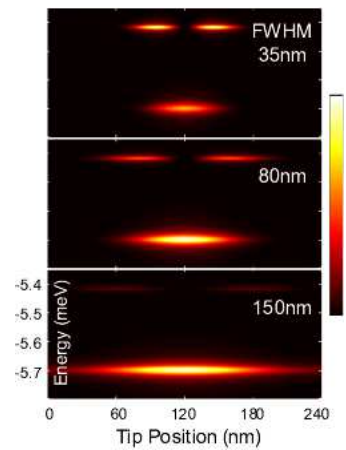}
 \caption{(color online).  PL spectral line-scans calculated at three different spatial resolutions, at the temperature of 4 K.}\vspace{-1 cm}
 \label{fig4}\end{center}
 \end{figure}
The influence of spatial resolution  on the PL spectral line-scans
is shown in Fig.\ 4. This figure displays the line-scans calculated
at $T= 2 K$ for three different spatial resolutions (indicated in
the figure). Lowering spatial resolution, the signal from the dark
state tend to disappear owing to cancellation effects in the matrix
element $o^{out}_{\alpha}$ governing the light-matter coupling.
Finally, we test the influence of random interface fluctuations with
a correlation length of the order of the exciton Bohr-radius \cite{zimm,zimm1}.
The total effective potential felt by excitons
is obtained adding to the dot potential used for all the previous
calculations a contribution modeled as a zero mean, Gauss
distributed and spatially correlated process with a correlation
length $\sigma=8$ nm and a  width of the energy distribution
$v_0=0.6$ meV (about the $10\%$ of the dot potential-well).
Fig.\ 5(a) displays a specific realization of this
potential.
Fig.\ 5(b) displays the energy-integrated PL image ($FWHM =40$ nm, $T= 2
K$) for the potential realization shown in panel 5(a). It evidences the efficiency of the dot capture and near-field
emission. Fig.\ 5(c) shows that the presence of the
symmetry-breaking disordered potential slightly increases the
linewidth of the second energy level, determining a  slight
equilibration of the emission lines.
\begin{figure} \begin{center}
\includegraphics[scale=0.6]{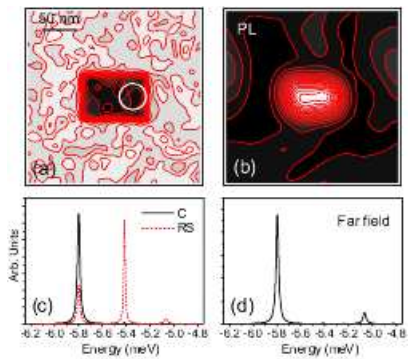}
 \caption{(color online). (a) Specific realization of the effective disordered potential used for the calculation of the PL images and spectra show. 
(b) PL energy-integrated image obtained  after uniform illumination of the sample at the barrier energy and  collecting locally ($FWHM =40$ nm, $T=4$ K). 
(c) PL spectra calculated centering the tip in the center of the dot and in the position indicated by a circle in the first panel ($FWHM =40$ nm, $T=4$ K).
(d)Far-field PL spectrum ($FWHM =40$ nm, $T=4$ K).}\vspace{-1 cm}
 \label{fig5}\end{center}
 \end{figure}
   Fig.\ 5(d) displays the
far-field PL spectrum. A very small peak (not present in the absence
of disorder) appears at the energy of the second level, which thus
remains mainly dark, demonstrating the robustness of dark-states
with respect to this kind of disorder. Results obtained for other
specific realizations of the random potential (here not shown),
don't display qualitative differences.

In conclusion, we have investigated the impact of the reduced
dark-states relaxation on the distribution of  electron-hole pairs
among the available energy levels after a far-field continuous wave
excitation. The calculated near-field luminescence properties of
these states depend critically on tip position, temperature and
spatial resolution and clearly indicate  the potentiality of
near-field PL  for adressing general questions regarding nanoscale
energy transfer in open nanosystems.


\begin{thebibliography}{100}
\bibitem{qdev} C.\ G.\ Smith, Rep. Prog. Phys.{\bf 59}, 235-282
(1996).
\bibitem{zimm} R.\ Zimmermann {\it et al}, Superlattices Microstruct. {\bf 17}, 439
(1995).

\bibitem{zimm1} R.\ Zimmermann and E.\ Runge, Phys. Status Solidi (a) {\bf 164}, 511 (1997).

\bibitem{hess} H. F.\ Hess, E.\ Betzig, T. D.\ Harris, L. N.\ Pfeiffer, and K. W.\ West, Science {\bf 264}, 1740 (1994).

\bibitem{guest} J.\ R.\ Guest, T.\ H\. Stievater, Gang Chen, E.\ A.\ Tabak, B.\ G\.
Orr, D.\ G.\  Steel, D.\ Gammon, and D.\ S.\ Katzer, Science
{\bf 293}, 2224 (2001).

\bibitem{matsuda} K.\ Matsuda {\it et al}, Phys. Rev. Lett. {\bf 91}, 177401 (2003).

\bibitem{runge} E.\ Runge, C.\ Lineau, Appl. Phys. B {\bf 84}, 103–110 (2006).

\bibitem{APLpippo} G.\ Pistone, S.\ Savasta, O.\ Di Stefano, and R.\ Girlanda, Appl. Phys. Lett. {\bf 84}, 2971 (2004).



\bibitem{mauritz} O.\ Mauritz, G.\ Goldoni, F.\ Rossi, and E.\ Molinari, Phys. Rev. Lett. {\bf 82}, 847 (1999).

\bibitem{APLomar} O.\ Di Stefano, S.\ Savasta, G.\ Martino, and R.\ Girlanda, Appl. Phys. Lett. {\bf 77}, 2804 (2000).

\bibitem{jap} O.\ Di Stefano, S.\ Savasta, and R.\ Girlanda, J. Appl. Phys. {\bf 91}, 2302 (2002).


\bibitem{PRBpippo} G.\ Pistone, S.\ Savasta, O.\ Di Stefano, and R.\ Girlanda, Phys. Rev. B {\bf 67}, 153305 (2003).








\bibitem{JPCMgiovanna} G.\ Martino, G.\ Pistone, S.\ Savasta, O.\ Di Stefano, and R.\ Girlanda, J. Phys.: Condens. Matter {\bf 18}, 2367 (2006).
\bibitem{prbdistef} O.\ Di Stefano, S.\ Savasta, G.\ Martino, and R.\ Girlanda, Phys.Rev B {\bf 62}, 11071 (2000).

\bibitem{APLHohenester} U.\ Hohenester, G.\ Goldoni, and E.\ Molinari, Appl. Phys. Lett. {\bf84}, 3963 (2004).

\bibitem{PRLHohenester} U.\ Hohenester, G.\ Goldoni, and E.\ Molinari, Phys. Rev. Lett. {\bf 95}, 216802 (2005).


\bibitem{Axt} V. M.\ Axt, K.\ Victor, and A.\ Stahl, Phys. Rev. B {\bf 53}, 7244 (1996).






\end{thebibliography}
\end{document}